\documentclass[11pt,twoside]{article}

\usepackage{asp2006}
\usepackage{epsf}

\newcommand{\hmpc}{h^{-1}{\rm Mpc}}

\begin{document}
\title{Missing Halo Baryons and Galactic Outflows}   
\author{Romeel Dav\'e}   
\affil{University of Arizona}    

 \begin{abstract} 
We present predictions for galactic halo baryon fractions from
cosmological hydrodynamic simulations with a well-constrained model
for galactic outflows.  Without outflows, halos contain roughly the
cosmic fraction of baryons, slightly lowered at high masses owing to
pressure support from hot gas.  The star formation efficiency is large
and increases monotonically to low masses, in disagreement with data.
With outflows, the baryon fraction is increasingly suppressed in halos
to lower masses.  A Milky Way-sized halo at $z=0$ has about 60\% of
the cosmic fraction of baryons, so ``missing" halo baryons have largely
been evacuated, rather than existing in some hidden form.  Large halos
($\ga 10^{13}M_\odot$) contain 85\% of their cosmic share of baryons,
which explains the mild missing baryon problem seen in clusters.
By comparing results at $z=3$ and $z=0$, we show that most of the baryon
removal occurs at early epochs in larger halos, while smaller halos
lose baryons more recently.  Star formation efficiency is maximized in
halos of $\sim 10^{13}M_\odot$, dropping significantly to lower masses,
which helps reconcile the sub-$L_*$ slope of the observed stellar and
halo mass functions.  These trends are predominantly driven by {\it
differential wind recycling}, namely, that wind material takes longer to
return to low-mass galaxies than high-mass galaxies.  The hot gas content
of halos is mostly unaffected by outflows, showing that outflows tend
to blow holes and escape rather than deposit their energy into halo gas.
\end{abstract}


\section{Introduction}   

Everywhere we look, baryons are missing.  Globally, observations only
account for around half the baryons today~\citep[e.g.][]{fuk04}.
Simulations indicate that the missing baryons are contained in
intergalactic gas at $10^5<T<10^7$~K, called the Warm-Hot Intergalactic
Medium~\citep[WHIM; e.g.][]{dav01}.  Initially, \ion{O}{vi} absorbers seen
in quasar spectra were thought to be collisionally-ionized tracers of WHIM
gas~\citep{tri00}.  But recent simulations and observations suggest that
most intergalactic \ion{O}{vi} is actually photo-ionized~\citep[and refs
therein]{opp08b}, and so the WHIM must be traced using higher ionization
lines~\citep[e.g. \ion{O}{vii};][]{nic05}.

Baryons are also missing on galactic scales.  Current hierarchical
structure formation models predict that baryons do not substantially
decouple from dark matter until well inside of halos, hence the
expectation is that halos should contain roughly the cosmic fraction
of baryons.  However, dynamical modeling of the Milky Way's disk and
halo reveals that it contains at best half of its cosmic share of
baryons in stars and cold gas~\citep{deh98,som01}; the same is true
of M31~\citep{kly02}.  This is the ``Missing Halo Baryons" problem.
Either a substantial portion of the $\sim L_*$ galaxies' halo
baryons are in some heretofore hidden form, or else there has been
a sizeable exodus of baryons into the intergalactic medium (IGM).
Many models have advocated the former possibility, suggesting by
analogy with the WHIM that the missing halo gas is in some warm-hot
component~\citep[e.g.][]{fuk06,som06}, or cold clouds~\citep{mal04}.
However, there is little observational evidence for massive coronae of
hot gas around typical spirals~\citep{ben00,wan07}, and furthermore if
all halos contained such coronae, this would substantially overpredict
the soft X-ray background~\citep{pen99,wu01}.  Meanwhile, the mass in cold
clumps as traced by high velocity clouds is unlikely to be sizeable even
in optimistic scenarios~\citep{bli99,som06}.  In contrast, \citet{sil03}
developed an analytic model that argued for a substantial fraction of
baryons being removed by galactic outflows.  In fact, Silk presciently
predicted that such outflows could solve a surprisingly wide range of
current dilemmas in galaxy formation, which our simulations have largely
confirmed.

Clusters also seem to have a missing baryon problem.  In a recent
census by \citet{gon07}, baryons make up $\approx 13$\% of the mass
of the cluster, while the latest WMAP-5 results~\citep{hin08} favor a
cosmic mean value of $\approx 17$\%.  This is a small but persistent
discrepancy which could have implications for using clusters as probes
of precision cosmology.

In these proceedings we examine the baryonic content of galaxy
halos in cosmological hydrodynamic simulations with and without
galactic outflows.  The outflow model implemented in our simulations
is unique in that it matches detailed properties of a wide range of
outflow-related observables, including IGM enrichment at various
epochs~\citep{opp06,opp08b,opp09}, the galaxy mass-metallicity
relation~\citep{fin08}, early galaxy luminosity functions~\citep{dav06},
and intragroup gas enrichment and entropy~\citep{dav08}.  Even though
our modeling of outflows is parameterized and heuristic, these successes
suggest that it plausibly moves mass, metals, and energy on large scales
in a manner consistent with the real Universe.

\section{Halo Baryon Fractions}   

\begin{figure}
\vskip -0.5in
\setlength{\epsfxsize}{1.1\textwidth}
\centerline{\epsfbox{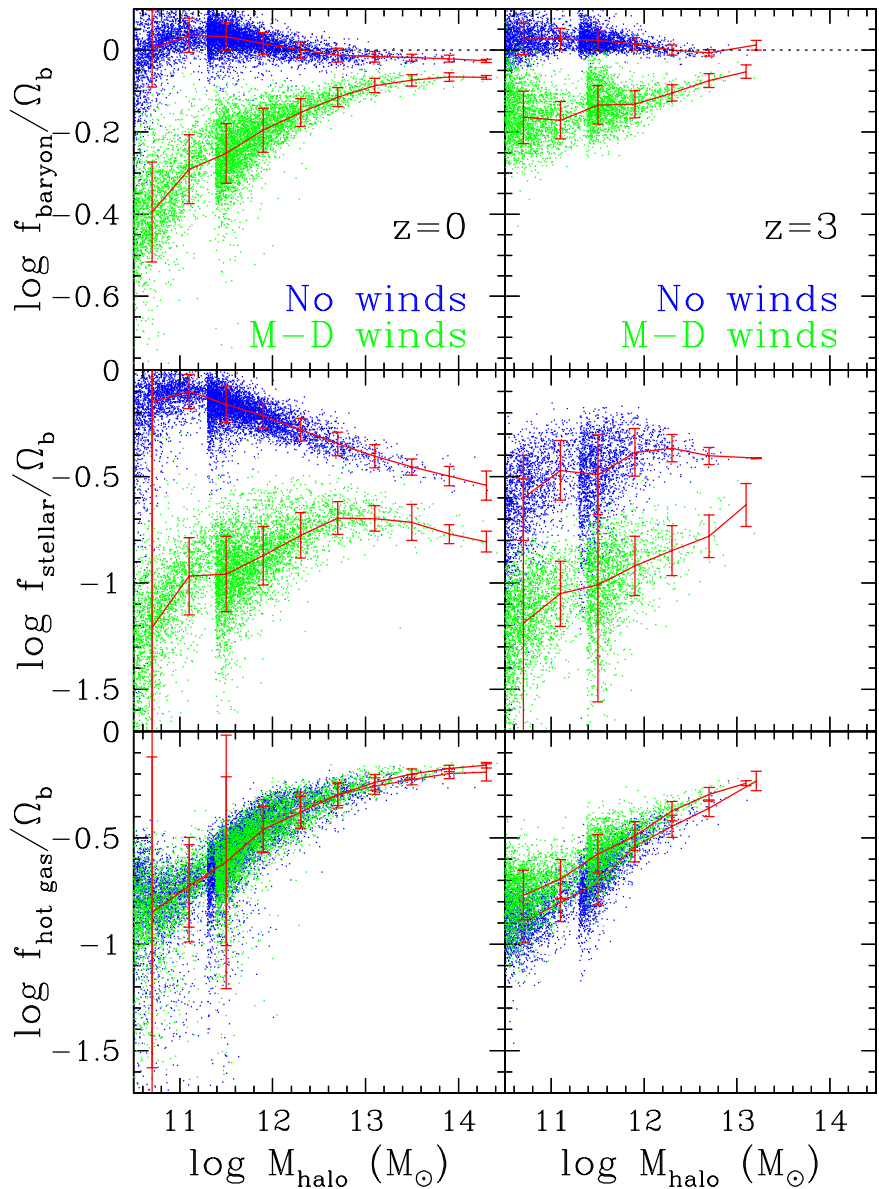}}
\vskip -0.5in
\caption{{\it Top:} Baryon fraction (relative to cosmic mean) as a function
of halo mass, for simulations at $z=0$ (left panels) and $z=3$ (right),
without winds (blue, upper points) and with momentum-driven winds (green, lower
points).  Dotted line shows the cosmic baryon fraction assumed in our runs.
{\it Middle:} Stellar fractions.
{\it Bottom:} Hot gas ($T>10^{4.5}$K) fractions.
In each panel, the red line shows a running median with $1\sigma$
dispersion, and the two groups of points for each model show results from
our $32\hmpc$ and $64\hmpc$ volumes.  Outflows drive out substantial
amounts of baryons particularly from low-mass systems, which helps
regulate star formation to observed level.  In large systems most of the
mass loss has already occured by $z=3$, while in small systems the mass
loss occurs later owing to differential wind recycling.  Hot gas fractions
are a strong function of halo mass, but are not sensitive to outflows.
A Milky Way-sized halo today contains about 60\% its cosmic share of
baryons, of which half is in the form of hot gas, one-third in cool gas,
and one-sixth in stars.
}
\label{fig:halofcomp}
\end{figure}

We run simulations using our modified version of Gadget-2~\citep{spr05},
as described in \citet{opp08a}.  The runs here employ $2\times 256^3$
particles in volumes of $32\hmpc$ and $64\hmpc$, with our WMAP3-concordant
{\it d-series} cosmology~\citep[see][]{opp08b}.  We identify halos
using a spherical overdensity algorithm~\citep[see][]{ker05},  and
consider only ``resolved" halos with masses $M_h>128(m_d+m_g)$,
where $m_d=(1.6,12.7)\times 10^8 M_\odot$ and $m_g=(0.34,2.72)\times
10^8 M_\odot$ are the dark matter and gas particles masses for the
$(32,64)\hmpc$ volumes, respectively.  We run two versions of these
simulations: One with no winds, and one with our favored momentum-driven
(M-D) wind model that matches a range of data as mentioned above.
Outflows are implemented in a probabalistic Monte Carlo fashion by giving
kicks to gas particles; details are in \citet{opp08a}.

The fraction of baryons within halos are shown in the top panel
of Figure~\ref{fig:halofcomp}.  In the no-wind case (blue points,
upper swath), halos of Milky Way's size and below have, as expected,
roughly their cosmic share of baryons (shown by the dotted line at
$\Omega_b/\Omega_m=0.044/0.25=0.176$).  In detail, adiabatic contraction
causes the baryon fraction to typically be slightly greater than the
cosmic mean.  This shows that without any strong feedback, baryons do
indeed mostly trace dark matter on halo scales in $\la L_*$ galaxies.

In more massive halos ($M_h\ga 10^{13}M_\odot$), the increasing
predominance of hot halo gas~\citep[][see also bottom panels of
Figure~\ref{fig:halofcomp}]{ker08} causes pressure support that
pushes baryons farther out than the dark matter, and the baryon
content is lowered.  At the highest masses probed by $z=0$ ($M_h\sim
10^{14}M_\odot$), this reduction reaches 6\%, which is not trivial but
still insufficient to explain observed cluster baryon fractions.

Now we consider the M-D wind runs, shown by the lower (green) swath of
points.  Outflows have a dramatic impact, increasingly so to smaller halo
masses.  Already by $z=3$, typical halos ($M_h\sim 10^{11-12}M_\odot$)
have had their baryon fraction reduced by $20-30$\% compared to the
cosmic mean.  At $z=3$ there is only a mild trend with halo mass, but by
$z=0$ the trend becomes much stronger:  Milky Way-sized halos ($M_h\sim
10^{12}M_\odot$) have lost roughly 40\% of their baryons, while the
smallest halos we probe at $M_h\sim 10^{10.5}M_\odot$ have lost 60\%.
This indicates that {\it missing halo baryons have mostly been ejected
by outflows.}

Outflows have a noticeable impact on large halos as well.  \citet{dav08}
showed that outflows add substantial entropy on poor group scales, and
this translates into increased pressure support over the no-wind case.
With outflows, the most massive ($M_h\ga 10^{13}M_\odot$) halos now
contain only 85\% of their cosmic share of baryons, and the trend is
essentially flat with halo mass.  This value is in better agreement with
observational estimates \citep{vik06,gon07}.

The falling baryon fraction in low-mass halos is a consequence
of wind recycling in our outflow model, i.e. the re-accretion of
previously ejected wind material.  In \citet{opp08a}, we found that
small galaxies push their winds out farther relative to large galaxies
(owing primarily to the fact that they reside in less dense surroundings),
and as a result the time for winds to be re-accreted onto small systems
is longer.  At early epochs, recycling is relatively unimportant, because
there hasn't been sufficient time to permit substantial re-accretion.
The weak trend seen at $z=3$ owes to the fact that our mass loss rate
scales inversely with velocity dispersion in our momentum-driven wind
scalings.  By $z=0$, however, the longer recycling times in smaller systems
results in much more cumulative mass loss from small halos relative to
larger halos.  Hence {\it differential wind recycling is the dominant
driver in establishing the baryon fraction trends with halo mass.}  As
we will see next, it is also critical for regulating the star formation
efficiency in smaller systems.

\section{Stellar Baryon Fractions}

We now separate baryons within each halo into three phases: Stars,
cool gas ($T<10^{4.5}$K), and hot gas ($T>10^{4.5}$K).  Star-forming
gas is included in cool gas.  The middle and bottom panels of
Figure~\ref{fig:halofcomp} show the baryon fractions in stars and hot
gas, respectively; the remainder is in cool gas (not shown).  In this
section we examine stellar baryon fractions.

With no outflows, there is a serious overcooling problem:  Star formation
is far too efficient.  In sub-$L_*$ halos, the fraction of baryons
in stars approaches 90\%, while observed small galaxies have stellar
fractions under 10\%.  Moreover, the trend is wrong; with no winds,
smaller galaxies have a higher baryon fraction in stars, while the
shallow slope of the stellar mass function below $L_*$ relative to the
halo mass function indicates that the stellar mass fraction should be lower
in small halos.  At group scales, the stellar baryon fraction is
down to 25\% of the cosmic mean, but that is still too high compared to
data as pointed out in \citet{dav08}.  Of course, these results are not
surprising, as it is well known that outflows, particularly at early
epochs, are required to suppress overcooling~\citep[e.g.][]{dav06}.

With outflows, the situation improves dramatically.  Stellar fractions are
reduced to a maximum of $\approx 20$\% for $\sim 10^{13}M_\odot$ halos,
falling to either smaller or larger halos.  A Milky Way-sized halo now has
a $12\%$ stellar fraction (relative to $\Omega_b$), which is in general
agreement with observations of comparable disk galaxies~\citep{ham07}.
The stellar fraction drops towards lower halo masses, suggesting a flatter
stellar mass function compared to the halo mass function, as observed.
The slope of the $z=0$ fit (red line) is approximately $d\log f_*/d\log
M_h\approx 0.5$ for $M_h<5\times 10^{12}M_\odot$, which when combined
with the halo mass function slope of $\approx -2$, yields a stellar mass
function slope of $\approx -1.5$.  This is in general agreement with
observations of the sub-$L_*$ stellar mass function slope~\citep{bal08},
though still too steep; in fact, the faint-end slope turns out to be
even shallower, as will be shown in a forthcoming paper.  Note that the
stellar fraction slope at $z=3$ is less steep, reflecting the shallower
slope in the overall halo baryon fraction (top right panel).  Our models
naturally yield a flattening of the faint end slope with time.

The qualitative trend of the star formation efficiency having a maximum
at some mass and dropping fairly rapidly to low masses, along with the
faint end slope evolution, is a direct consequence of differential wind
recycling.  Without it, the star formation efficiency would continue
to increase to small masses, as in the no-wind case.  Hence models of
galaxy formation must not only include outflows, but must also track
the {\it dynamics} of wind material on its journey through the IGM to
properly capture its impact on galaxy evolution.

\section{Hot Gas Baryon Fractions}

Turning to the hot gas ($T>30,000$~K) baryon fractions (bottom panels of
Figure~\ref{fig:halofcomp}), there is a strong trend of hot gas fraction
increasing with halo mass, which is mostly independent of redshift and
whether or not outflows are included.  In the no-wind case, the hot
gas fraction exceeds the stellar fraction (which comprises the vast
majority of condensed baryons) at $z=0$ for $M_h\ga 10^{12.5}M_\odot$.
This transition occurs at a somewhat larger mass than found in \citet{ker08},
likely because our runs include metal-line cooling.

With outflows, the fraction of halo gas in hot form is more substantial,
exceeding the stellar fraction in the outflow case at all masses
probed here.  This is due to the suppression of star formation,
rather than an increase in the amount of hot gas.  Relative to the
baryon fraction, the hot gas fraction is about one-third of all halo
baryons at the smallest masses probed, and increases to three-fourths
or more for $M_h\ga 10^{13}M_\odot$.  This indicates that there is
still a substantial reservoir of baryons in a difficult-to-detect phase
within typical galaxies, and indeed X-ray observations do indicate a
warm-hot corona around a nearby spiral galaxy~\citep{ped06}.  However,
the amount of hot gas we predict is still a factor of two to three smaller than
in models that place the majority of missing halo baryons into this phase.
This roughly translates into up to an order of magnitude reduction in the
predicted X-ray flux~\citep[as compared to e.g.][]{ben00}.  Hence the
typical non-detection of X-ray halos around spiral galaxies should not
be surprising, but deeper observations with future X-ray telescopes 
should uncover this phase more ubiquitously.

It is noteworthy that our outflows do not significantly increase
the amount of hot gas into halos (as seen by comparing to the
no-wind case).  Winds in our models tend to escape halos without
depositing a significant amount of energy along the way via shocks.
This is contrary to typical assumptions in analytic or semi-analytic
models of outflows~\citep[e.g.][]{dek86}, and simply reflects the
fact that in realistic three-dimensional models outflows prefer to
blow holes rather than share their energy with ambient gas.  This is
also consistent with X-ray observations of hot gas around galaxies
that indicate they are radiating a small fraction of the supernova
energy input~\citep[e.g.][]{wan07}.  Another factor is that outflows
are typically quite enriched, so that the gas cooling times are short,
and hence even if outflow material is heated it condenses out quickly.
At low-$z$ this results in a ``halo fountain" that may be the origin of
compact high velocity clouds~\citep{wan07,opp08a}.

\section{Summary}   

We have examined halo baryon fraction in cosmological hydrodynamic
simulations, comparing models with and without galactic outflows.
Our outflow model is heuristic, but is well-constrained to match a
variety of galaxy and IGM observations at a range of epochs.  We find
that these same outflows may help resolve some puzzles regarding the
baryonic content of galactic halos.

For low mass halos, outflows remove a significant portion of halo baryons
by $z=0$.  The fraction of ejected baryons increases sharply to lower
masses, so that $10^{12}M_\odot$ halos have lost 40\% of their baryons,
and $10^{10.5}M_\odot$ halos 60\%.  The trend is not as pronounced at
$z=3$, showing that the longer wind recycling times in small galaxies
(i.e. differential wind recycling) plays a critical role in suppressing
the baryon fractions in small halos between $z=3\rightarrow 0$.  This may
explain why the faint-end slope of the luminosity function becomes
shallower with time.  By today, this produces a peak in star formation
efficiency at $\sim 10^{12.5-13}M_\odot$, above and below which the efficiency
falls.  Meanwhile, hot gas fractions are mostly unaffected by outflows,
showing that outflows do not deposit much of their energy into halo gas,
instead preferring to blow holes and escape.  A Milky Way-sized halo
today has about 60\% of its cosmic share of baryons, half of which are
in hot gas, one-third in cool gas, and the remainder in stars.

For high mass halos ($M_h\ga 10^{13}M_\odot$), there is a mild
suppression of baryon fractions owing to larger pressure support
from an increasingly substantial hot gaseous halo.  Without winds,
the suppression is fairly small ($\approx 5\%$).  In our outflow run,
conversely, the suppression is much more substantial, $\approx 15\%$,
independent of mass, which agrees better with data.  This demonstrates
that outflows impact even massive halos today.  The key is that most of
the ejection occurs at early epochs, when those halos were much smaller.

Our implementation of outflows is primarily constrained to match IGM
enrichment at $z\sim 2-4$~\citep[e.g.][]{opp06}.  The fact that this
model naturally yields observationally-consistent results for halo baryon
fractions at $z\sim 0$ is highly encouraging, and represents another
significant success for our outflow model based on momentum-driven wind
scalings.  It may be that we are approaching a heuristic understanding
of how outflows operate on extragalactic scales, even though the
detailed mechanisms by which outflows are driven out of galaxies
remain unclear.


\acknowledgements 

RD thanks Shardha Jogee and the organizing committee for an excellent
meeting.  The simulations were run on UA's SGI 
cluster.


\end{document}